# An Efficient Analytical Solution to Thwart DDoS Attacks in Public Domain

B. B. Gupta
Department of Electronics and Computer Engineering
Indian Institute of Technology, Roorkee-247667, India
bbgupta@ieee.org

R. C. Joshi
Department of Electronics and Computer Engineering
Indian Institute of Technology, Roorkee-247667, India

Manoj Misra
Department of Electronics and Computer Engineering
Indian Institute of Technology, Roorkee-247667, India

## ABSTRACT
In this paper, an analytical model for DDoS attacks detection is proposed, in which propagation of abrupt traffic changes inside public domain is monitored to detect a wide range of DDoS attacks. Although, various statistical measures can be used to construct profile of the traffic normally seen in the network to identify anomalies whenever traffic goes out of profile, we have selected volume and flow measure. Consideration of varying tolerance factors make proposed detection system scalable to the varying network conditions and attack loads in real time. NS-2 network simulator on Linux platform is used as simulation testbed. Simulation results show that our proposed solution gives a drastic improvement in terms of detection rate and false positive rate. However, the mammoth volume generated by DDoS attacks pose the biggest challenge in terms of memory and computational overheads as far as monitoring and analysis of traffic at single point connecting victim is concerned. To address this problem, a distributed cooperative technique is proposed that distributes memory and computational overheads to all edge routers for detecting a wide range of DDoS attacks at early stage.

## 1. INTRODUCTION
Breaches in network security represent a growing problem to businesses and institutions, costing them billions of dollars every year. According to statistics given by CERT [1], a mere 171 vulnerabilities were reported in 1995 that boomed to 7236 in 2007. Already, the number for the same has gone up to 6058 until the third quarter of 2008. Apart from these, a large number of vulnerabilities go unreported every year. In particular, Denial-of-Service (DoS) attacks are a major threat to the Internet. DDoS attacks are commonly characterized as events where legitimate users or organizations are deprived of certain services like web, e-mail or network connectivity that they normally expect to have [2]. Therefore, as given by Weiler [3] they attempt: (1) To inhibit legitimate network traffic by flooding the network with useless traffic. (2) To deny access to a service by disrupting connections between two parties. (3) To block the access of a particular individual to a service. (4) To disrupt the specific system or service itself. Series of DDoS attacks that shut down some high profile websites have demonstrated the severe consequences of these attacks [4]. As per computer crime and security survey conducted by FBI/CSI in the United States for the year 2004 [5], these attacks are the second most widely detected outsider attack types in computer networks immediately after virus infections. A computer crime and security survey conducted in Australia for the year 2004 [6] shows similar results.

In this paper, we have proposed an analytical model, which concentrates on detection of a wide range of DDoS attacks, e.g. high rate disruptive, diluted low rate degrading and varied rate, by monitoring propagation of abrupt traffic changes inside public domain. Although, various statistical measures can be used to identify anomalous behavior of the system, we have selected volume and flow measure. Consideration of varying tolerance factors make proposed detection system scalable to the varying network conditions and attack loads in real time. Internet type topologies used for simulations are generated using Transit-Stub model of GT-ITM topology generator [7]. NS-2 [8] on Linux platform is used as simulation testbed. Simulation results demonstrate that proposed scheme inflicts an extremely high detection rate with low false positive rate. However, the mammoth volume generated by DDoS attacks pose the biggest challenge in terms of memory and computational overheads as far as monitoring and analysis of traffic at single point connecting victim is concerned. These overheads make DDoS solution itself vulnerable against these attacks. To address this problem, we present a distributed cooperative technique that distributes memory and computational overheads to all edge routers for detecting variety of DDoS attacks at early stage. Two types of detectors are used in this approach: local detector and central detector.

The remainder of this paper is organized as follows: section 2 contains DDoS overview, 3 points out related work, section 4 presents analytical model for DDoS attack detection, section 5 describes our proposed approach in detail, section 6 contains experimental design and performance analysis and section 7 describes proposed distributed cooperative techniques for DDoS attack detection. Finally, Section 8 concludes the paper and outlines future work.

## 2. DDOS OVERVIEW

DDoS is basically a resource overloading problem. The resource can be bandwidth, memory, CPU cycles, file descriptors and buffers etc. The attackers bombard scare resource either by flood of packets or a single logic packet, which can activate a series of processes to exhaust the limited resource [9]. In the Figure 1, simplified DDoS attack scenario is illustrated. The figure shows that attacker uses three zombie machines to generate mammoth volume of malicious traffic to flood the victim over the Internet thus rendering legitimate user unable to access the services. Extremely sophisticated, user friendly, automated and powerful DDoS toolkits are available for attacking the victim, so expertise is not necessarily required that attract naive users to perform DDoS attacks [10].

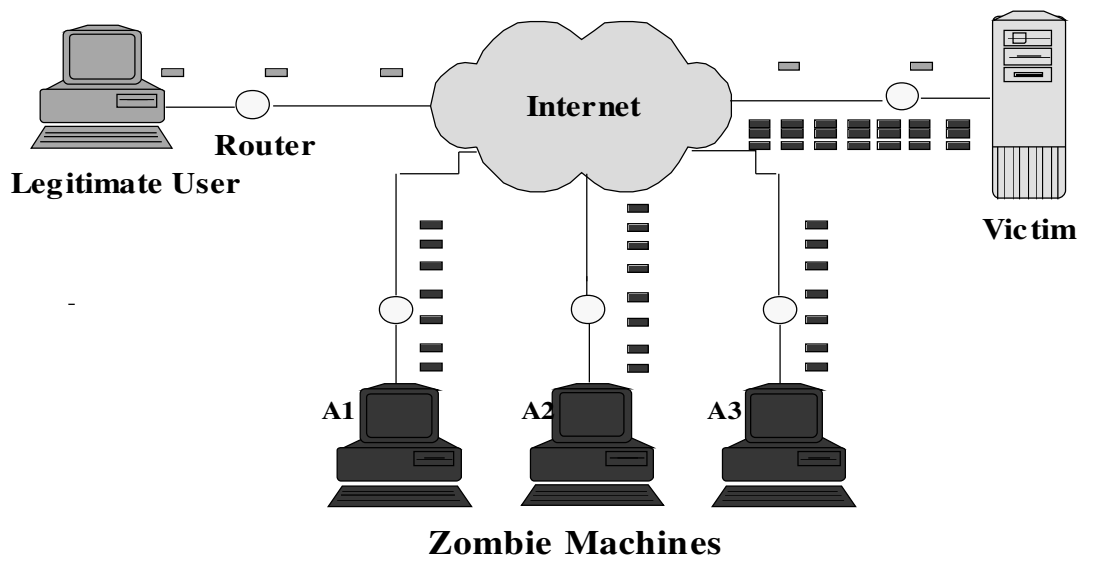

**Figure 1. Illustration of the DDoS attack scenario**

## 2.1 Attack: Modus Operandi

DDoS attack is carried out from multiple sources to aim at a single target, in several phases. Figure 2 shows a hierarchical model of a DDoS attack. In order to launch a DDoS attack, the attacker first scan millions of machines for vulnerable services and other weakness on the Internet through high-bandwidth, always-on connections that permit penetrations. The discovered vulnerabilities are then exploited to gain access and plant malicious codes on these machines so called handlers, or masters. After being installed the malicious scripts, these infected machines can repeat the same procedure to recruit more machines so called zombies or slaves. These all exploited machines used as attack army, are collectively called bots and the attack network is known as botnet in the hacker's community. Then the communication channels between the attacker and the masters and between the masters and slaves are established. These control channels are designed to be secret to public [11]. Staying behind the scenes of attack, the real attacker sends a command to the masters to initiate a coordinated attack. When the masters receive the command, they transfer it to the slaves under their control. Upon receiving attack commands, the zombies or slaves begin the attack on the victim [12].

The real attacker is trying to hide himself from detection, for example, by providing spoofed IP addresses [11, 13].

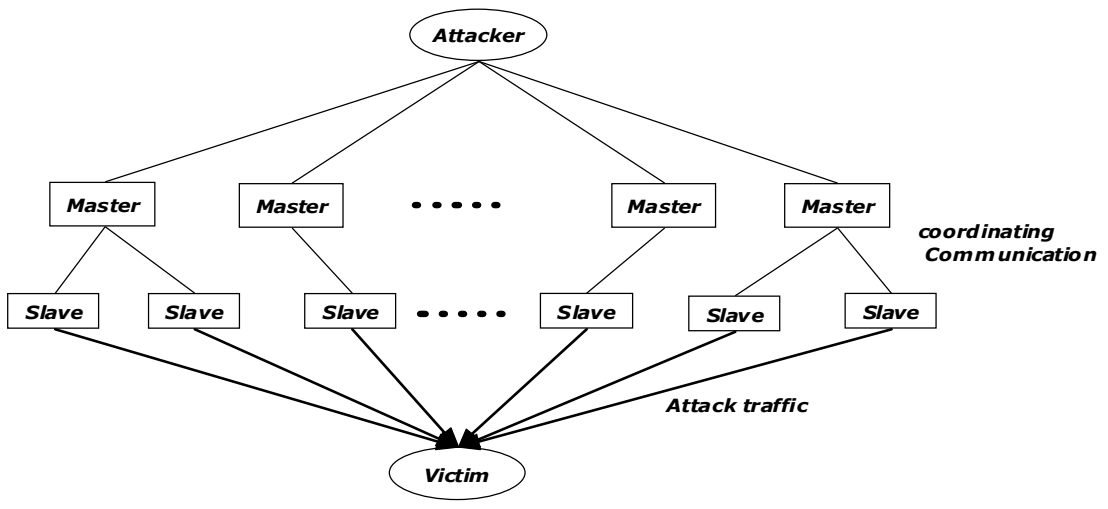

**Figure 2. A hierarchical model of a DDoS attack**

## 3. RELATED WORK

Various reviews have been given in [10, 13, 14] on DDoS attacks and defense methods. Attack detection aims to detect an ongoing attack and to discriminate malicious traffic from legitimate traffic.

Detection can be performed using database of known signatures, by recognizing anomalies in system behaviors or using third party. Signature based approach [15, 16] employs a priori knowledge of attack signatures. The signatures are manually constructed by security experts analyzing previous attacks and used to match with incoming traffic to detect intrusions. SNORT [15] and Bro [16] are the two widely used signature based detection approaches.

Anomaly detection [17-20] relies on detecting behaviors that are abnormal with respect to some normal standard. Detecting DDoS attacks involves first knowing normal behavior of our system and then to find deviations from that behavior. Mirkovic et al. [17] proposed D-WARD defense system that does DDoS attack detection at source, based on the idea that DDoS attacks should be stopped as close to the source as possible. Cheng et al. [18] proposed to use spectral analysis to identify DoS attack flows. Lee and Stolfo [19] used data mining techniques to discover patterns of system features that describe program and user behavior and implement a classifier that can recognize anomalies and intrusions. Bencsath et al. [20] have given a volume based approach, in which incoming traffic is monitored continuously and dangerous traffic intensity rises are detected. This approach is suitable for detecting high rate attacks, but ineffective to detect low rate degrading attacks.

Mechanisms that deploy third-party detection do not handle the detection process themselves, but rely on an external third party that signals the occurrence of the attack [13].

# 4. ANALYTICAL MODEL DESCRIPTION

This section describes an analytical model we have constructed to represent general form of proposed detection system. Detecting DDoS attacks involve first knowing normal profile of the system and then to find deviations from this normal profile. Whenever incoming traffic goes out of the normal profile, anomalous system behavior is identified. Our approach detects flooding DDoS attacks by the constant monitoring the propagation of abrupt traffic changes inside the public domain. Various statistical measures e.g. volume, flow, entropy, ratio etc are available for profile generation.

Let $M$ and $F$ are random vectors compose of $m$ measures $m_1, m_2, \ldots, m_m$ used for attacks detection and $n$ flows $f_1, f_2, \ldots, f_n$ containing the incoming traffic to the server, respectively: $M=(m_1, m_2, \ldots, m_m)$, $F=(f1, f2, \ldots, fn)$. where $f_i=(m_1^i, m_2^i, \ldots, m_m^i)$ is $i^{th}$ flow. Consider a random process $\{m_j^i(t), t=\omega\Delta, \omega\in N\}$, where $\Delta$ is a constant time interval, $N$ is the set of positive integers, and for each $t$, $m_j^i(t)$ is a random variables. $1\le\omega\le l$, $l$ is the number of time intervals. Here $m_j^i(t)$ represents the value of $m_j$ in flow $i$ in $\{t-\Delta, t\}$ time duration. These relations can be written in matrix form as follows:

$$Z(t)=\begin{pmatrix} m_1^1(t) & m_1^i(t)\cdots & m_1^n \\ m_j^1(t) & m_j^i(t)\cdots & m_j^n \\ \vdots & \vdots & \vdots \\ m_m^1(t) & m_m^i(t)\cdots & m_m^n(t) \end{pmatrix} \quad (1)$$

here, $Z(t)$ contains values of different measures used in $\{t-\Delta, t\}$. $m_j(t)$ represent total value of $j^{th}$ measure during $\{t-\Delta, t\}$ time. $m_j(t)$ can be calculated as follows:

$$m_j(t)=m_j^1(t)+m_j^2(t)+\ldots\ldots+m_j^i(t)+\ldots\ldots+m_j^n(t) \quad (2)$$

where $1\le i\le n$, $n$ is the number of flows. $1\le j\le m$, $m$ is the number of measures. Normal traffic value of $j^{th}$ measures can be calculated using following equation:

$$\overset{*}{m}_j(t)=\frac{1}{l}\sum_{\omega=1}^{l} m_j(t=\omega\Delta)$$

(3)

Vector $A$ can be used to represent normal traffic measures value: $A=(\overset{*}{m}_1(t), \overset{*}{m}_2(t), \ldots, \overset{*}{m}_m(t))$. To detect the attack, the value of $j^{th}$ traffic measure $m_j(t)$ is calculated in time window $\Delta$ continuously; whenever there is appreciable deviation from $\overset{*}{m}_j(t)$, anomalous behaviors could be determined. Depending on the measures selected to use or network conditions, following event is defined to determine anomalous system behaviors:

$$m_j(t)-\overset{*}{m}_j(t)>\xi_j$$

(4)

where $\xi_j$ represent value of the threshold for $j^{th}$ measure. $\xi_j$ can be set as follows:

$$\xi_j=r_j*\sigma_j \quad (5)$$

where $\sigma_j$ represent value of standard deviation for $j^{th}$ measure. $r_j$ represent value of tolerance factor to calculate the threshold for $j^{th}$ measure. Effectiveness of an anomaly based detection system highly depends on accuracy of threshold value settings. Inaccurate threshold values cause a large number of false positives and false negatives. Therefore, various simulations are performed using different value of tolerance factors. The choice of tolerance factors varies for different network conditions. Values of tolerance factors also depend on the composition of the normal traffic and the desired degree of the ability to control a DDoS attack. Then, trade-off between detection and false positive rate provides guidelines for selecting value of tolerance factor $r_j$ for $j^{th}$ traffic measure for a particular simulation environment.

# 5. PROPOSED APPROACH

Detection system is part of access router or can belong to separate unit that interact with access router to detect attack traffic. Detecting DDoS attacks involve first knowing normal traffic model of our system and then to find deviations from this normal traffic model. Our approach detects flooding DDoS attacks by monitoring the propagation of abrupt traffic changes inside the network. Analytical model describe in the section 4 is used here. Although, various statistical measures can be used to capture anomalous behavior of the system, we have selected volume and flow measure.

Let $X_n^*(t)$ be the normal traffic, indicating total bytes arriving at a target machine in $\Delta$ time duration. Assume that the target machine is intruded by DDoS attacks at time $t_a$. Generally, target may not overwhelm immediately at $t_a$. Assume attacker has attack traffic rate such that it overwhelms target completely at time $t_b$, so target is unable to provide any service to its customers. Time duration ($t_a$, $t_b$) is known as transition period of attack. A good detection approach must have detection time $t_d<t_b$, so the target may be avoided being overwhelmed completely. Let $X_{in}(t)$ be the traffic during transition period ($t_a$, $t_b$), then we can express $X_{in}(t)$ as follows:

$$X_{in}(t)=X_n^*(t)+\hat{\hat{X}}(t), \quad (8)$$

where $\hat{\hat{X}}(t)$ is the component of the attack traffic. $X_{in}(t)$ - $X_n^*(t)$ using equation (8) can be used for detection purpose. To set normal profile, consider a random process $\{X(t), t=\omega\Delta, \omega\in N\}$, where $\Delta$ is a constant time interval, $N$ is the set of positive integers, and for each $t$, $X(t)$ is a random variable. $1\le\omega\le l$, $l$ is the number of time intervals. Here $X(t)$ represents the total traffic volume in $\{t-\Delta, t\}$ time interval. $X(t)$ is calculated during time interval $\{t-\Delta, t\}$ as follows:

$X(t) = \sum_{i=1}^{N_f} n_i, i = 1, 2....N_f$ . Here, $n_i$ represents total number of bytes arrivals for a flow i in {$t$–Δ, $t$} and $N_f$ represent total number of flows. We take average of $X(t)$ and designate it as $X_n^*(t)$ normal traffic Volume. Similarly, value of flow measure is calculated and designates that as $F_n^*(t)$. Here total bytes, not packets are used to calculate volume measure, because it provides more accuracy, as different flows can contain packets of different sizes. To detect the attack, the value of traffic measure $X_{in}(t)$ and flow measure $F_{in}(t)$ is calculated in time window Δ continuously; whenever there is appreciable deviation from $X_n^*(t)$ and $F_n^*(t)$, various types of attacks are detected using algorithm 1 as given in figure 3. Threshold values $\xi_{th}$ and $\varsigma_{th}$ are set as follows:

$$\xi_{th} = r * \sigma_V \qquad (9)$$

$$\varsigma_{th} = r * \sigma_F \qquad (10)$$

where $\sigma_V$ , $\sigma_F$ represents value of standard deviation for volume measure and flow measure, respectively.

---

***Algorithm 1: DDoS attacks Detection Algorithm***

***Input:*** *$X_n^*(t)$ : Normal traffic Volume measure*

*$F_n^*(t)$ : Normal traffic Flow measure*

*$\xi_{th}$ : Threshold value for Volume measure*

*$\varsigma_{th}$ : Threshold value for Flow measure.*

***Output:*** *DDoS attack alert generation.*

***Procedure:***

***01:*** *Consider a random process { $X_{in}(t)$, $F_{in}(t)$, $t = \omega\Delta$ , $\omega \in N$ }, where Δ is a constant time interval, N is the set of positive integers, and for each t , $X_{in}(t)$ and $F_{in}(t)$ are random variables. $1 \le \omega \le l$, l is the number of time intervals. Here, $X_{in}(t)$ represents value of volume measure and $F_{in}(t)$ represents value of flow measure in time duration {t–Δ, t}.*

***02: If*** *(( $X_{in}(t) - X_n^*(t) > \xi_{th}$ ) || ( $F_{in}(t) - F_n^*(t) > \varsigma_{th}$ ))* ***Then***
*Attack pattern detected.*
*DDoS attack alert is generated.*

***03: Else If*** *(( $X_{in}(t) - X_n^*(t) < \xi_{th}$ ) && ( $F_{in}(t) - F_n^*(t) < \varsigma_{th}$ ))*
***Then*** *System is attack free.*
*Attack alert is not generated.*

---

**Figure 3. Algorithm for DDoS attacks Detection**

We have taken same value of tolerance factor for both volume and flow measure. $r \in I$ , represent value of tolerance factor, where *I* is a set of integers.

## 6. EXPERIMENT DESIGN AND ANALYSIS

We tested and evaluated proposed approach with monitoring data to confirm its effectiveness to detect variety of DDoS attacks.

### 6.1 Simulation Environment

The simulation is carried out using NS-2 [8] network simulator. At present, the Internet can be viewed as a collection of interconnected routing domains, which are groups of nodes under a common administration that share routing information. A primary characteristic of these domains is routing locality, in which the path between any two nodes in a domain remains entirely within the domain. Thus, each routing domain in the Internet can be classified as either a stub or a transit domain [21]. A domain is a stub domain if the path connecting nodes $N_1$ and $N_2$ passes through that domain and if either $N_1$ or $N_2$ is located in that domain. Transit domains do not have this restriction. The purpose of transit domains is to interconnect stub domains efficiently. Therefore, real-world Internet type topologies generated using Transit-Stub model of GT-ITM [7] topology generator is used to test our proposed scheme, where transit domains are treated as different Internet Service Provider. Topology contains four transit domains. Total four hundred legitimate client machines are used to generate background traffic. Total zombie machines range between ten and hundred to generate attack traffic. Transit domain four contains the server machine to be attacked. A short scale simulation topology is shown in figure 4. The legitimate clients are TCP agents that request files of size 1 Mbps each with request inter-arrival times drawn from a Poisson distribution. The attackers are modeled by UDP agents. A UDP connection is used instead of a TCP one because in a practical attack flow, the attacker would normally never follow the basic rules of TCP, i.e. waiting for ACK packets before the next window of outstanding packets can be sent, etc. The rate of attack traffic range between 0.1 and 3.5 Mbps per attack daemon. The size of monitoring window affects the number of attack alerts raised. In our experiments, the monitoring time window was set to 200 ms, as the typical domestic Internet RTT is around 100 ms and the average global Internet RTT is 140 ms [22].

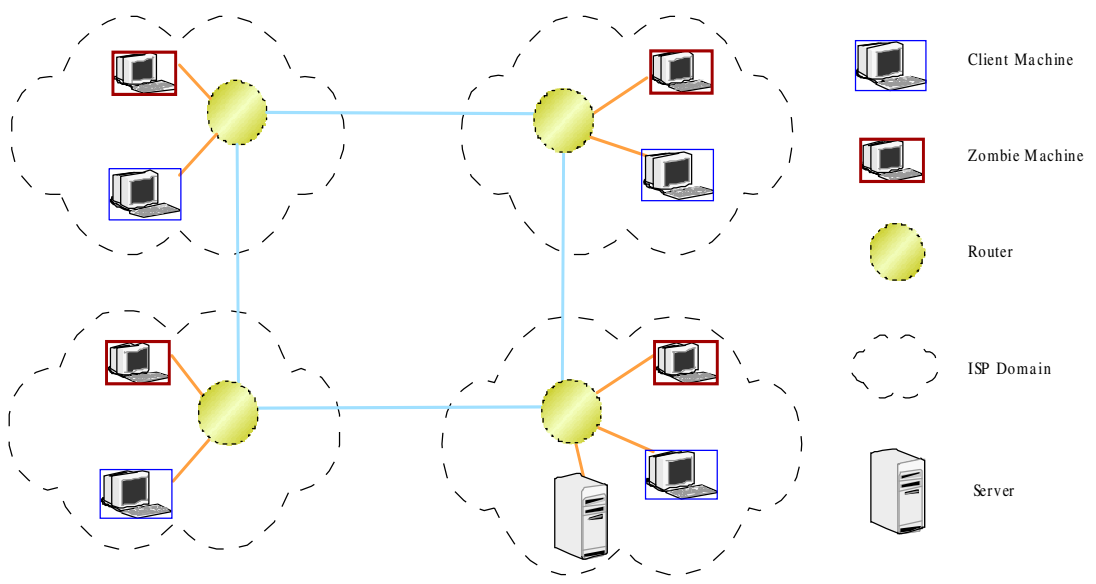

**Figure 4. A short scale simulation topology**

Total false positive alarms are minimum, using this value of monitoring window.

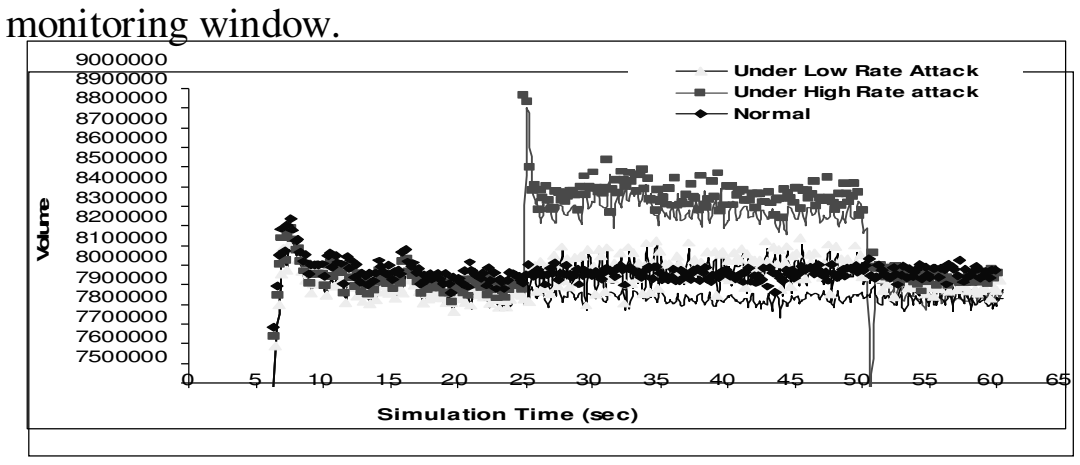

**Figure 5. Temporal variation of volume measure when system is in normal condition, under low rate DDoS attack, and under high rate DDoS attack**

False positive alarm number increases steadily with increasing monitoring window size. The simulations are repeated and different attack scenarios are implemented by varying total number of zombie machines and at different attack strengths. Figure 5 shows temporal variation of volume measures when system is in normal condition, under low rate DDoS attack and under high rate DDoS attack. DDoS attacks start at 25th second and end at 50th second. Here total 400 client machines are used to send TCP traffic. Low rate attack is performed using 100 zombie machines with mean rate 0.1Mbps per attacker. To perform high rate attack 100 zombie machines are used with mean rate 3Mbps per attacker. As shown in figure 5, it is clear that low rate attacks are nearly undetectable when using only volume as statistical measure.

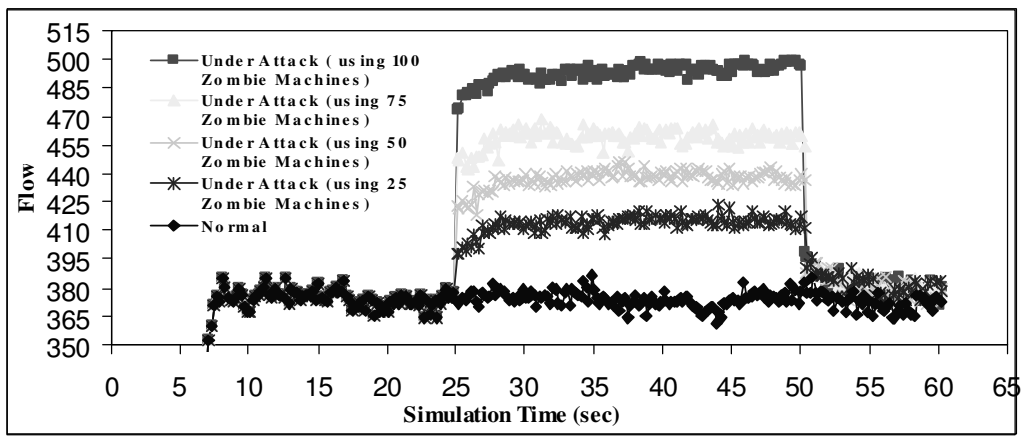

**Figure 6. Temporal variation of Flow measure when system is in normal condition, and under DDoS attack using 25, 50, 75 and 100 zombie machines**

For detection of low rate DDoS attack correctly with low false positive rate, flow measure is also considered along with volume measure. Figure 6 shows temporal variation of flow measure when system is in normal condition and under DDoS attack using 25, 50, 75 and 100 zombie machines.

It is clear from the figure 5 and figure 6, that low rate DDoS attacks perform using large number of zombie machines are also easily detected by taking both flow and volume measure simultaneously.

## 6.2 Performance Evaluation Metrics

We have used three metrics to evaluate performance of our proposed DDoS detection approach, namely, detection rate ($R_d$), false positive alarm rate ($R_{fp}$), and receiver operating characteristic (ROC). The detection rate ($R_d$) is the measure of percentage of attacks detected among all actual attacks performed. The detection rate ($R_d$) is defined as follows:

$$R_d = d/n \quad (11)$$

Where d is the number of DDoS attacks detected and n is the total number of actual attacks generated during the simulation. The false positive alarm rate ($R_{fp}$) is the measure of percentage of false positives among all normal traffic events defined as follows:

$$R_{fp} = f/m \quad (12)$$

where f is the number of false positive alarm raised by attack detection mechanism, and n is the total number of normal traffic flow events during the simulation. The ROC curve is used to evaluate tradeoff between detection rate and false positive rate.

## 6.3 Simulation Results and Discussion

Figure 7 illustrates the variation of the detection and false positive rate with respect to different value of detection tolerance factor r, when DDoS attack is performed using packets of size 512 bytes and 1024 bytes. Detection rate is close to 99% when r <=6 and false positive alarm rate is <=3% when r >=6. Above result demonstrates that detection rate is very high with low false positive rate when r=6. The ROC curve in figure 8 also shows same results. Therefore, value of r is taken 6 in our approach. Value of r can vary for different network conditions and correct value can be selected by drawing tradeoff between detection and false positive rate.

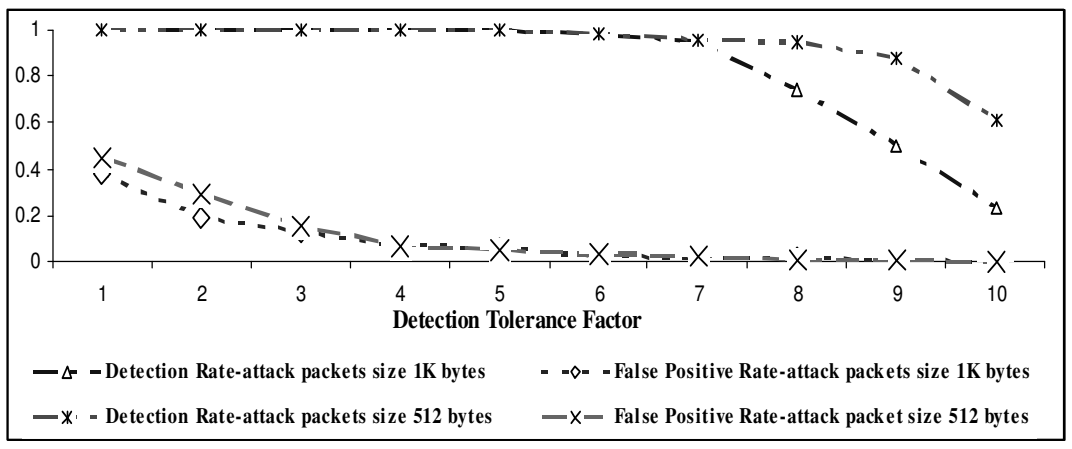

**Figure 7. Effect of detection tolerance factor on the detection and false positive rate when attack is perform using packets of size 512 bytes and 1024 bytes**

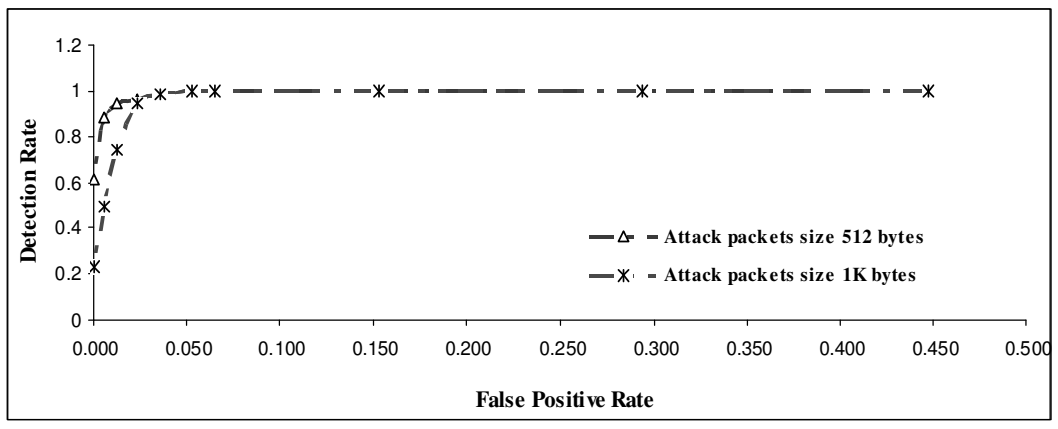

**Figure 8. ROC curve showing the trade-off between the detection rate and false positive rate of DDoS attacks when attack is perform using packets of size 512 bytes and 1024 bytes**

# 7. DISTRIBUTED COOPERATIVE TECHNIQUE

DDoS attacks are best detected near the victim's site i.e. at bottleneck router, as maximum attack traffic converges at this point. However, the mammoth volume generated by DDoS attacks pose the biggest challenge in terms of memory and computational overheads as far as monitoring and analysis of traffic at single point connecting victim is concerned. These overheads make DDoS solution itself vulnerable against these attacks. To address this problem, we present a distributed cooperative technique that distributes memory and computational overheads to all edge routers for detecting a wide range of DDoS attacks at early stage. Two types of detectors are used in our approach: local detector and central detector.

## 7.1 Local Detector

Local detectors are deployed on edge routers of the domain. These detectors are used to check flow statistics i.e. volume and flow measure, during time $\Delta$ duration. Whenever there is appreciable deviation from normal profile, a Suspicious Alarm (SA) is reported to central detector.

## 7.2 Central Detector

Central detector is deployed on router to which protected server is attached. Central detector detects attacks not only by passively listening for SA from local detectors, but also by actively sending queries to the local detectors to confirm alarms. Central detector also checks flow statistics for attack detection during constant time window. Central detector passively waits for suspicious alarms to arrive from edge routers, and whenever it crosses threshold, DDoS attacks confirmation alarm is generated.

All the calculation regarding volume and flow metrics are done at edge routers. These information plus attacks alert plus active flow list are sent to access router then based on the alert receive, access router decides about attack. Simultaneously access router itself checks the statistics and then compares with the threshold and generates attack alert alarm. Therefore, no need of computational and storage requirement at access router as edge router do all the calculation. For flow measure calculation, flows list from all edge routers are added and common flows are deleted. Then total flows are counted from that list.

# 8. CONCLUSION AND FUTURE WORK

In this paper, we analytically detect a wide range of DDoS attacks i.e. high rate disruptive, diluted low rate degrading and varied rate by monitoring the propagation of abrupt traffic changes. Simulation results are very promising by using volume and flow measure to capture anomalous behavior of the system. Proposed approach have limitation in terms of memory and computational overheads as monitoring and analysis of traffic is performed at single point i.e. bottleneck router. To address this problem, a distributed cooperative technique is proposed that distributes memory and computational overheads to all edge routers for detecting a wide range of DDoS attacks at early stage. Although simulation results are promising, but in future work we plan to validate our approach with real datasets.

# 9. ACKNOWLEDGMENTS

The authors gratefully acknowledge the financial support of the Ministry of Human Resource Development (MHRD), Government of India for partial work reported in the paper.